\documentclass{PoS}

\PoS{PoS(LAT2005)035}
\usepackage{psfrag}

\title{OPE Analysis of QCD Potential and Determination of $\Lambda_{\overline{\rm MS}}$ }

\ShortTitle{OPE Analysis of QCD Potential and Determination of $\Lambda_{\overline{\rm MS}}$ }

\author{\speaker{Yukinari Sumino}\\
        Tohoku University\\
        E-mail: \email{sumino@tuhep.phys.tohoku.ac.jp}}

\abstract{
We analyze the static QCD potential in the distance region
$0.1~{\rm fm} \simlt r \simlt 1$~fm.
We combine most recent  
lattice computations and 
perturbative computations of the potential, in the framework of
operator-product expansion (OPE).
We determine simultaneously the non-perturbative contribution to the potential,
$\delta E_{\rm US}(r) $, and the relation between the
lattice scale (Sommer scale) and $\LMS$
in the quenched approximation.
We find that 
(1) large part of the short-distance linear potential belongs to
the perturbative Wilson coefficient,
(2) $\delta E_{\rm US}(r) =0$ is disfavored, and
(3) $r_0\, \Lambda_{\overline{\rm MS}}^{\mbox{\tiny{3-loop}}}
=0.574\pm 0.042 $.
It provides a new method for precise determination of 
$r_0\, \Lambda_{\overline{\rm MS}}^{\mbox{\tiny{3-loop}}}$.
}

\FullConference{XXIIIrd International Symposium on Lattice Field Theory\\
                 25-30 July 2005\\
                 Trinity College, Dublin, Ireland}

\newcommand \bra[1]{\left< {#1} \,\right\vert}
\newcommand \ket[1]{\left\vert\, {#1} \, \right>}

\newcommand{\bea}{\begin{eqnarray}}
\newcommand{\eea}{\end{eqnarray}}
\newcommand{\simgt}{\hbox{ \raise3pt\hbox to 0pt{$>$}\raise-3pt\hbox{$\sim$} }}
\newcommand{\simlt}{\hbox{ \raise3pt\hbox to 0pt{$<$}\raise-3pt\hbox{$\sim$} }}

\newcommand{\LQ}{\Lambda_{\rm QCD}}
\newcommand{\LMS}{\Lambda_{\overline{\rm MS}}}

\begin{document}

\section{Introduction}

For some years, the static QCD potential $V_{\rm QCD}(r)$
has been successfully computed
in lattice simulations.
For instance, the distance $r$ 
of the potential computed in some of most recent lattice computations in the quenched approximation
ranges from about 0.1~fm to beyond 2~fm.
Computed results by different groups, when superimposed with one another, 
come more or less on a single curve
with fairly good accuracy, showing very good stability of the lattice predictions
(see Fig.~\ref{comp-VS-lat}(a) below).
In particular, they show that numerically the QCD potential tends to a linear 
potential smoothly at large distances, in accord with the
confinement picture of the quarks.

In this paper, we analyze the potential in the distance region
which is smaller than (but not too small as compared to) 
the typical hadron scale $\LQ^{-1}$.
More specifically, we consider the region
$0.1~{\rm fm} \simlt r \simlt 0.5$--1~fm if expressed in physical units.
This is the region that is relevant to 
spectroscopy of the heavy quarkonia, such as bottomonium and charmonium.
In this region, it is known empirically that the static potential
can be approximated well by a Coulomb + linear form.

In this distance range, accuracy of the perturbative predictions for the
static potential improved drastically around year 1998
\cite{Hoang:1998nz}.
It was proposed to subtract IR renormalon contained in the $r$-independent
(constant) part of the potential, e.g.\ by
computing the total energy $2m_{\rm pole}+V_{\rm QCD}(r)$ after reexpressing 
$m_{\rm pole}$
in terms of the $\overline{\rm MS}$ mass or by computing the force.
Then,
we find dramatic improvement in convergence of the perturbative
series, as well as much more stable perturbative predictions against
scale variation.
Moreover, once we obtain much more accurate predictions,
good agreement with phenomenological potentials
and with lattice computations of the potential has been observed
in the distance range $0.1~{\rm fm} \simlt r \simlt 0.5$~fm;
by now, this has been confirmed by several groups \cite{Sumino:2001eh,Necco:2001xg,Pineda:2002se}.

In this article we review our recent work \cite{Sumino:2005cq}, in which we analyze
the QCD potential in the framework of operator-product expansion
(OPE) \cite{Brambilla:1999xf}.
We assemble most recent developments of the perturbative
predictions and lattice computations.
Using OPE, one can separate perturbative contributions
from non-perturbative contributions
in an unambiguous manner.
In this way, we are able to determine
(1) the size of non-perturbative contributions to the potential, and
(2) the relation between $\LMS$ and lattice scale (Sommer scale).
In practical applications, the latter determination would be particularly
important, since it can be used to determine
$\alpha_S(M_Z)$ with high accuracy in near future, when lattice simulations 
incorporating dynamical quarks with realistic masses will become
available.

\section{OPE of the QCD potential}

The OPE of the static QCD potential was developed around year 1999 
\cite{Brambilla:1999xf},
within the framework of potential-NRQCD effective field theory \cite{Pineda:1997bj}.
Conceptually it is close to OPEs of other physical quantities, with which one
may be more familiar.
When there is a small parameter as compared to the typical hadron size
(in our case the distance $r \ll \LQ^{-1}$ 
between static quark and antiquark),
one constructs a series of operators as an expansion in $r$.
Each operator has a Wilson coefficient, which is perturbatively
computable, whereas a non-perturbative contribution is contained
in the matrix element of the operator.
The OPE of the QCD potential reads
\bea
V_{\rm QCD}(r) = V_S(r) + \delta E_{\rm US}(r)
\eea
with
\bea
\delta E_{\rm US}  \propto \alpha_S \int_0^\infty dt \,
\, e^{-i \, 
t \, \{ V_O(r) - V_S(r) \} } \,
\bra{0} \vec{r}\cdot\vec{E}^a(t) \,
\varphi(t,0)_{ab} \,
\vec{r}\cdot\vec{E}^b(0) \ket{0}
+ {\cal O}(r^3) .
\eea

Here, $V_S(r)$ denotes the Wilson coefficient corresponding to the
leading UV contribution to the potential (whose operator is the identity operator).
$V_S(r)$ is referred to as ``singlet potential,'' is perturbatively computable,
and has a Coulombic behavior (with logarithmic corrections)
$\sim -1/|r \log r|$ at short-distances.
We note that $V_S(r)$ is different from the perturbative expansion of
$V_{\rm QCD}(r)$;
all IR contributions (IR divergences and IR renormalons)
have been subtracted and absorbed into a non-perturbative contribution,
so that the perturbative expansion of $V_S(r)$ is more convergent
than that of $V_{\rm QCD}(r)$.
See \cite{Sumino:2005cq} for the precise definition of $V_S(r)$.

On the other hand, $\delta E_{\rm US}(r)$ denotes the
leading non-perturbative contribution.
It is given as an integral over time of an operator matrix element.
The operator is local in space but non-local in time.
The behavior of $\delta E_{\rm US}(r)$ at small-distances is known
exactly to be ${\cal O}(\LQ^4 r^3)$ when $r$ is very small
(when $V_O-V_S \approx C_A \alpha_S/r \gg \LQ$ holds).
It is, however, doubious whether this strong condition is met in the distance
region of our interest, $0.1~{\rm fm} \simlt r \simlt 1$~fm.
In this distance region, where it is more likely that $V_O-V_S \sim \LQ$, the behavior of
$\delta E_{\rm US}(r)$ cannot be predicted model-independently.
According to some models it is
estimated to be ${\cal O}(\LQ^3 r^2)$.
In any case, in the following analysis, we will use only the fact that
$\delta E_{\rm US}(r) \to 0$ as $r \to 0$, which is true because at very
small distances $\delta E_{\rm US}(r)={\cal O}(\LQ^4 r^3)$.

Intuitively one can interpret in the following way.
It is the contributions of gluons, whose wavelengths are smaller than
$r$, that generates the Coulomb singularity of the potential
as $r \to 0$.
On the other hand, non-perturbative 
contributions can be regarded as contributions of
gluons whose wavelengths are larger than $r$.
Such gluons see only the total charge of the system, hence, as $r \to 0$,
they decouple from the color-singlet system.
Therefore,
their contributions vanish as $r \to 0$.

\section{Comparison of {\boldmath $V_S(r)$} and recent lattice computations}

\begin{figure}
\hspace*{-17mm}
\includegraphics[width=.63\textwidth]{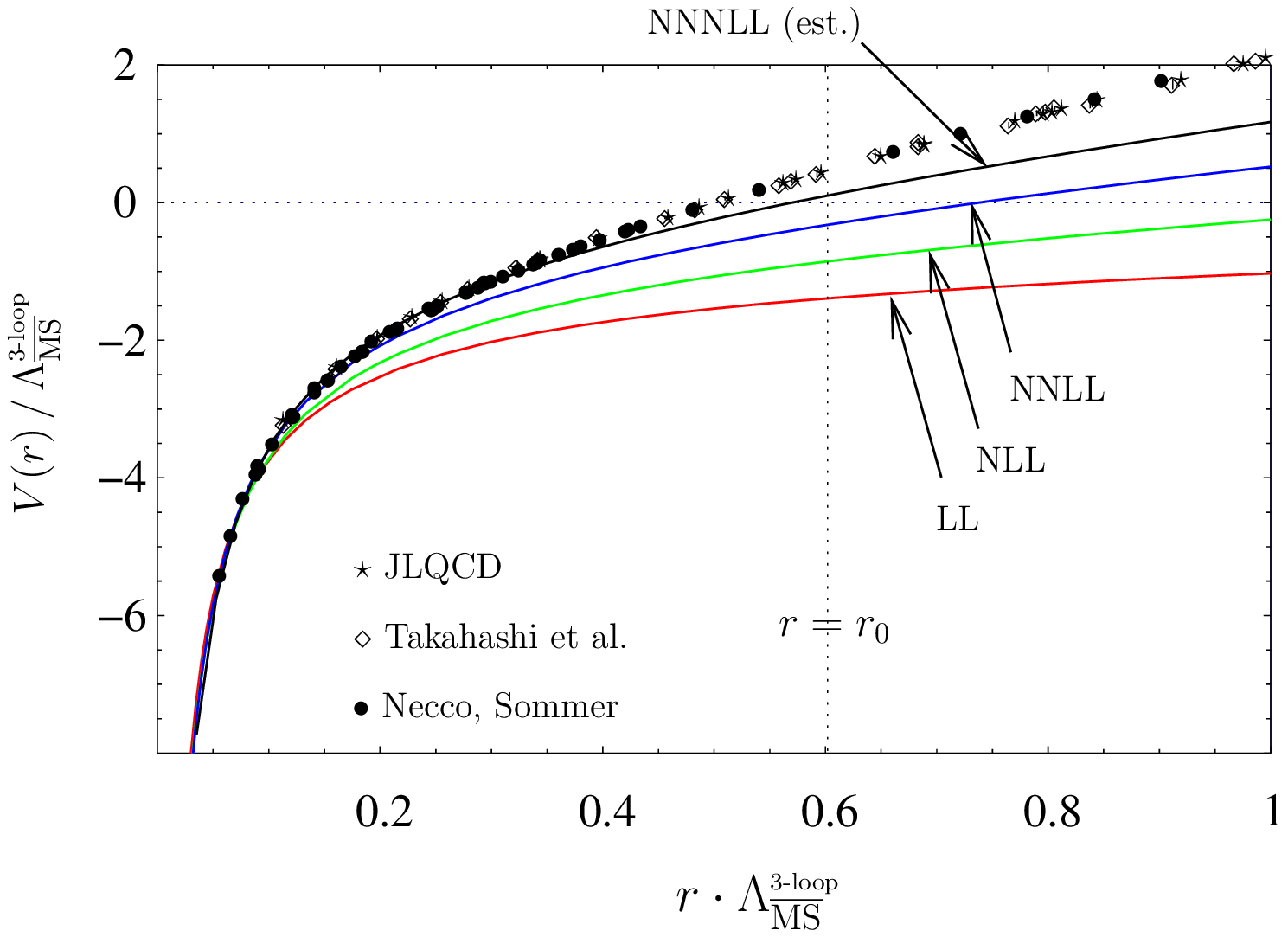}
\psfrag{LL}{\footnotesize  LL}
\psfrag{NLL}{\hspace{2mm}\footnotesize  NLL}
\psfrag{NNLL}{\footnotesize NNLL}
\psfrag{NNNLL}{\hspace{0mm}\footnotesize NNNLL}
\psfrag{Large-beta0}{\hspace{-3mm}\footnotesize large-$\beta_0$ appr.}
\psfrag{r0}{\hspace{-6mm} $r=r_0$}
\psfrag{delEUS}{\hspace{-13mm} $[V_{\rm latt}(r)-V_S(r)]/
\Lambda_{\overline{\rm MS}}^{\mbox{\scriptsize 3-loop}}$}
\psfrag{r}
{\hspace{-3mm}
$r \cdot \Lambda_{\overline{\rm MS}}^{\mbox{\scriptsize 3-loop}}$}
\hspace*{5mm}
\includegraphics[width=.4\textwidth]{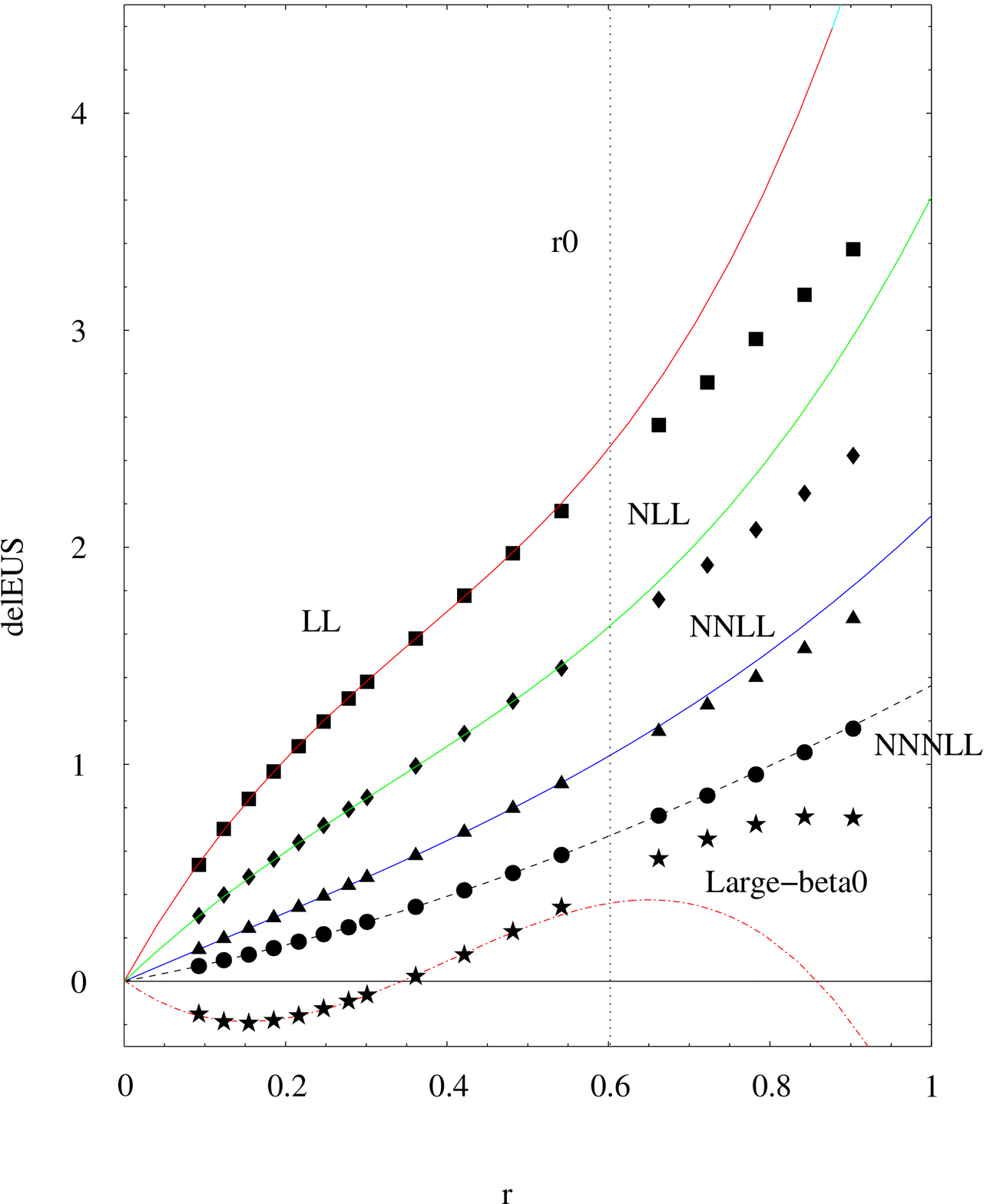}\\
\hspace{44mm}(a)
\hspace{68mm}(b)
\caption{
(a) Comparison of the Wilson coefficient $V_S(r)$ and lattice data.
(b) $V_{\rm latt}(r)-V_S(r)$ vs.\ $r$.
We set $r_0\, 
\Lambda_{\overline{\rm MS}}^{\mbox{\tiny{3-loop}}}
=0.602$ only in these figures.
}
\label{comp-VS-lat}
\end{figure}

In order to make a numerical cross chek of the prediction of OPE,
we first compare the perturbative predictions of the leading Wilson
coefficient $V_S(r)$ and
recent lattice results.
In Fig.~\ref{comp-VS-lat}(a) we plot lattice data of the QCD potential from 
three different
groups \cite{Necco:2001xg,Takahashi:2002bw}.
In the same figure, the perturbative predictions of $V_S(r)$ are
plotted up to the
next-to-next-to-next-to-leading logarithmic order (NNNLL).
As a salient feature, we see that the perturbative predictions
follow the lattice data up to larger distances as we increase the order.

Following input parameters were chosen in depicting this figure.
First, we note that in order to make a comparison between $V_S(r)$
and lattice data,
the relation between $\LMS$ and lattice scale
(we take the Sommer scale $r_0$)  is needed.
This is because, a priori
each lattice data set
is given in units of $r_0$, whereas perturbative predictions
are given in units of $\LMS$.
Thus, conversion of units is needed to compare them in common units.
Here, only in this section, we use the central value of the conventionally
known relation $r_0\, 
\Lambda_{\overline{\rm MS}}^{\mbox{\tiny{3-loop}}}
=0.602 \pm 0.048$ \cite{Capitani:1998mq}.
Other inputs are:
$\alpha_S(Q)=0.2$, $n_l=0$;
at NNNLL, the 3-loop non-logarithmic term
$\bar{a}_3$ is not known yet, only some estimates exist,
hence, we used
Pineda's estimate for $\bar{a}_3$ \cite{Pineda:2002se}
and varied it within a range including
other estimates as a part of our estimates of uncertainties;
we chose C+L scheme in computing $V_S(r)$.

To verify the prediction of OPE, a closer inspection is needed.
Data points in Fig.~\ref{comp-VS-lat}(b) 
show the differences of $V_S(r)$ and the lattice data \cite{Necco:2001xg},
where the vertical scale is magnified as compared to Fig.~\ref{comp-VS-lat}(a).
At LL, one can still see a Coulombic contribution in the difference, but
as the order is increased, a Coulombic contribution vanishes and
they tend to be regular at the origin.
Solid lines in the same figure show fits of the data points
by cubic polynomials,
using data points at 
$r  \Lambda_{\overline{\rm MS}}^{\mbox{\tiny 3-loop}}
<0.5$.
One sees that the solid lines approximate the data points 
up to larger distances as the order increases,
which indicate that these data points become regular around
the origin as the order increases.
These features support the OPE prediction, that the non-perturbative
contribution 
$\delta E_{\rm US}(r)$ vanishes as $r \to 0$.
We note that such behavior can be observed only when we take the
difference between $V_S(r)$ and the high quality lattice data, since
in Fig.~\ref{comp-VS-lat}(a), 
perturbative predictions become so steep toward the origin that
any steep curve tends to go through the lattice data points.

\section{Determination of  $\delta E_{\rm US}(r)$ and
$r_0\, \Lambda_{\overline{\rm MS}}^{\mbox{\tiny{3-loop}}}$
}

In the light of the observation in the previous section, we
perform a simultaneous fit to determine the non-perturbative contribution
$\delta E_{\rm US}(r)$
and the relation between lattice scale and $\LMS$,
$x \equiv r_0\, 
\Lambda_{\overline{\rm MS}}^{\mbox{\tiny{3-loop}}}
$.
(In the previous section we chose a specific value for $x$ determined
from another source, but in this
section we determine its value using the QCD potential alone.)

Let us explain why we have a high sensitivity to the value of $x$.
It constitutes the essence of our analysis in this article.
We compare the lattice data for the potential and perturbative prediciton
of $V_S(r)$ in some {\it common units}.
Here, we choose to compare in units of $\LMS$, hence we convert the
lattice data into these units by using any chosen value of $x = r_0\, 
\Lambda_{\overline{\rm MS}}^{\mbox{\tiny{3-loop}}}
$.
Let us write
\bea
V_{\rm latt}(r;x)-V_S(r)
= [ V_{\rm latt}(r;x_{\rm true})-
V_S(r)]~+~[
V_{\rm latt}(r;x)-
V_{\rm latt}(r;x_{\rm true})] .
\eea
It is a trivial equality, where $x_{\rm true}$ denotes the true value of $x$.
The first bracket $[...]$ on the right-hand-side coincides with the
non-perturbative contribution $\delta E_{\rm US}(r)$, hence
it vanishes as $r \to 0$.
On the other hand, the second bracket on the right-hand-side takes
a ``Coulomb''+linear form as long as $x \neq x_{\rm true}$.
(Here, ``Coulomb'' includes logarithmic corrections at short-distances.)
It follows from the fact that $V_{\rm latt}(r;x)$ takes a ``Coulomb''+linear form
and that $[
V_{\rm latt}(r;x)-
V_{\rm latt}(r;x_{\rm true})]$ is merely the difference of these functions
after rescaling $V_{\rm latt}(r;x)$ in the 
vertical and horizontal directions.
Hence, the second term contains a Coulombic singularity as $r \to 0$ unless 
$x = x_{\rm true}$.
It means that in order to find the true value of $x$,
we should adjust $x$ such that
the leading Coulombic behavior vanishes
in $V_{\rm latt}(r;x)-V_S(r)$
as $r \to 0$.
That the fit is determined by the leading singular behavior guarantees a
high sensitivity to $x$.

As for $\delta E_{\rm US}(r)$, we simply fit it by a quadratic polynomial using
the data points at $r  \Lambda_{\overline{\rm MS}}^{\mbox{\tiny 3-loop}}
<0.5$, since $\delta E_{\rm US}(r)$ is regular as $r \to 0$.\footnote{
We also tried a cubic fit, where the results of the fit did not change considerably.
}

\begin{figure}
\hspace*{40mm}
\psfrag{A1}{$A_1$} 
\psfrag{A2}{$A_2$} 
\psfrag{68pcCL}{\footnotesize 68\%} 
\psfrag{95pcCL}{\footnotesize 95\%} 
\includegraphics[width=.4\textwidth]{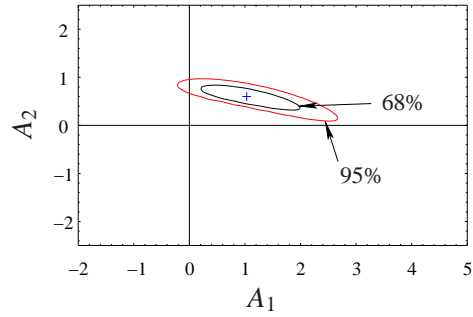}
\caption{
Allowed regions in the parameter space $(A_1,A_2)$ at
68\% and 95\% CL, when 
$
V_{\rm latt}(r;x)-
V_S(r)$
is fitted by $A_1 \cdot \LMS^2 r +A_2 \cdot \LMS^3 r^2$.
}
\label{fit-res}
\end{figure}

We show the results of the fit for $\delta E_{\rm US}(r)$
in Fig.~\ref{fit-res}.
The central value of the coefficient of the linear potential is
about 1 in units of 
$(\Lambda_{\overline{\rm MS}}^{\mbox{\tiny 3-loop}})^2$.
We note that the coefficient of the linear potential 
(string tension) determined from
the large distance behavior of lattice results is about 3.8 in the same units.
Hence, our result (central value) indicate that the non-perturbative contribution
contains only about one quarter of the string tension, as a 
component of the linear potential
in the short-distance region.
Three quarters of the string tension belongs to the perturbative 
contribution $V_S(r)$.
The present error is somewhat large, so it is still consistent at 95\% confidence level
that either {\it all} of the string tension is contained in  $V_S(r)$ or only
1/3 of it is contained in $V_S(r)$.

Another notable feature is that (in the schemes which we examined) 
$\delta E_{\rm US}(r) =0$ is strongly disfavored.
It is seen from Fig.~\ref{fit-res}, where the origin is excluded from the
allowed region.
It means that there is a non-vanishing non-perturbative contribution.
In fact, this is a new result, since in a previous similar study
by Pineda \cite{Pineda:2002se}, 
$\delta E_{\rm US}(r) $ contained a larger error and
was still consistent with zero.
The major difference of the analyses, which led to the improvement of
the bound in our analysis, is that we determined $x$ simultaneously
from the fit.

We also obtained
\bea
x = 0.574 \pm 0.042 .
\label{value-x}
\eea
It should be compared with the conventional values
determined by the Schr\"odinger functional method:
\bea
&&
x = 0.602 \pm 0.048
~~~~~\mbox{\cite{Capitani:1998mq}} ,
\\
&&
x = 0.586 \pm 0.048
~~~~~\mbox{\cite{Necco:2001xg}} .
\eea
The errors are of the same order of magnitude, and the
values are consistent with one another with respect to
the errors.
We emphasize that our determination uses a completely independent
method, and that the mutual consistency between these values
is quite non-trivial.
At the moment, the error of our $x$ in eq.~(\ref{value-x})
is dominated by errors of the lattice data of the potential.

\section{Conclusions}

We analyzed the static QCD potential in the distance region
$0.1~{\rm fm} \simlt r \simlt 1$~fm.
We combined the perturbative predictions up to NNNLL and 
the most accurate lattice computations, in the framework of
OPE.
In this way, we determined the non-perturbative contribution
$\delta E_{\rm US}(r) $ and 
$r_0\, 
\Lambda_{\overline{\rm MS}}^{\mbox{\tiny{3-loop}}}
$,
for $n_l =0$ (quenched approximation).
Our conclusions are as follows:
\begin{itemize}
\item
Most of the linear potential at $r<1$~fm
($\sim 3.8~\Lambda_{\overline{\rm MS}}^2
r$)
is included in perturbative prediction
of the Wilson coefficient {$V_S(r)$}.
This is a {\it scheme-independent} result.\footnote{
Scheme dependence enters at ${\cal O}(r^2)$, hence it does not
affect the linear potential.
}
\item
$\delta E_{\rm US}(r)=0 $ is disfavored.
This is a scheme-dependent result, but it holds in different schemes
which we examined (which we consider to be reasonable schemes).
\item
$r_0\, \Lambda_{\overline{\rm MS}}^{\mbox{\tiny{3-loop}}}
=0.574\pm 0.042 $.
It is consistent with the conventional values and the error is
of comparable size.
This provides a new method for its determination.

\end{itemize}


\begin{thebibliography}{99}
\bibitem{Hoang:1998nz}
A.~H.~Hoang, M.~C.~Smith, T.~Stelzer and S.~Willenbrock,
Phys.\ Rev.\ D {\bf 59}, 114014 (1999);
M.~Beneke,
Phys.\ Lett.\ B {\bf 434}, 115 (1998).

\bibitem{Sumino:2001eh}
Y.~Sumino,
Phys.\ Rev.\ D {\bf 65}, 054003 (2002);
S.~Recksiegel and Y.~Sumino,
Eur.\ Phys.\ J.\ C {\bf 31}, 187 (2003);
T.~Lee,
Phys.\ Rev.\ D {\bf 67}, 014020 (2003).

\bibitem{Necco:2001xg}
  S.~Necco and R.~Sommer,
  Nucl.\ Phys.\ B {\bf 622}, 328 (2002).

\bibitem{Pineda:2002se}
A.~Pineda,
J.\ Phys.\ G {\bf 29}, 371 (2003).

\bibitem{Sumino:2005cq}
  Y.~Sumino,
  arXiv:hep-ph/0505034.

\bibitem{Brambilla:1999xf}
N.~Brambilla, A.~Pineda, J.~Soto and A.~Vairo,
Nucl.\ Phys.\ B {\bf 566}, 275 (2000).

\bibitem{Pineda:1997bj}
  A.~Pineda and J.~Soto,
  Nucl.\ Phys.\ Proc.\ Suppl.\  {\bf 64}, 428 (1998).

\bibitem{Takahashi:2002bw}
  T.~T.~Takahashi, H.~Suganuma, Y.~Nemoto and H.~Matsufuru,
  Phys.\ Rev.\ D {\bf 65}, 114509 (2002);
  S.~Aoki {\it et al.}  [JLQCD Collaboration],
  Phys.\ Rev.\ D {\bf 68}, 054502 (2003).

\bibitem{Capitani:1998mq}
S.~Capitani, M.~L\"uscher, R.~Sommer and H.~Wittig  [ALPHA Collaboration],
Nucl.\ Phys.\ B {\bf 544}, 669 (1999); Erratum ibid.\ {\bf 582}, 762 (2000).



\end{thebibliography}
\end{document}